\documentclass[12pt]{article}
\usepackage{arxiv}
\usepackage[utf8]{inputenc} 
\usepackage[T1]{fontenc}    
\usepackage{hyperref}       
\usepackage{url}            
\usepackage{booktabs}       
\usepackage{amsfonts}       
\usepackage{nicefrac}       
\usepackage{microtype}      
\usepackage{lipsum}
\usepackage{amsmath}
\usepackage{cite}
\usepackage{graphicx,amssymb}
\usepackage{amsmath,amsfonts,amssymb}

\usepackage[usenames]{color}
\usepackage{algorithm}
\usepackage{algorithmic}
\usepackage{epstopdf}
\usepackage{multirow}
\usepackage{stfloats}
\usepackage{subfigure}

\usepackage{comment}
\specialcomment{warning}{\color{red}}{\color{black}}
\specialcomment{new}{\color{blue}}{\color{black}}
\specialcomment{remove}{}{} \excludecomment{remove}

\title{Decision Fusion in Space-Time Spreading aided Distributed MIMO WSNs}

\author{
Indrakshi Dey,
\thanks{Indrakshi Dey is with CONNECT Centre for Future Networks,
National University of Ireland,
 Maynooth, Ireland, Email: deyi@tcd.ie}
Hem Dutt Joshi, \thanks{Hem Dutt Joshi is with Department of ECE,
Thapar Institute of Engineering and Technology, Patiala, India, Email: hemdutt.joshi@thapar.edu}
and Nicola Marchetti \thanks{Nicola Marchetti is with CONNECT Centre for Future Networks
Trinity College Dublin, Ireland, Email: nicola.marchetti@tcd.ie}
}
\begin{document}
\maketitle

\begin{abstract}
In this letter, we propose space-time spreading (STS) of local sensor decisions before reporting them over a wireless multiple access channel (MAC), in order to achieve flexible balance between diversity and multiplexing gain as well as eliminate any chance of intrinsic interference inherent in MAC scenarios. Spreading of the sensor decisions using dispersion vectors exploits the benefits of multi-slot decision to improve low-complexity diversity gain and opportunistic throughput. On the other hand, at the receive side of the reporting channel, we formulate and compare optimum and sub-optimum fusion rules for arriving at a reliable conclusion. Simulation results demonstrate gain in performance with STS aided transmission from a minimum of 3 times to a maximum of 6 times over performance without STS.
\end{abstract}

\section{Introduction}

Channel aware decision fusion (DF) in a wireless sensor network (WSN) is the preferred way of exploiting the wireless medium as a multiple access channel (MAC) with the aim of attaining multiplexing gain \cite{banavar2012}. Implementing multiple antennas at the DFC has been recommended in \cite{dey2019throughput, Ciuonzo2012} to ameliorate fusion performance in deep fading and shadowing condition, leveraging diversity gain over a `virtual' multiple-input-multiple-output (MIMO) or massive MIMO (mMIMO) channel between the sensors and the decision fusion center (DFC). However, DF in MAC is corrupted with intrinsic interference resulting from inter-sensor-element interference (ISEI) and inter-sensor-channel interference (ISCI). Interference caused by partial overlap of multiple sensor signals in time results in ISEI and ISCI is caused by the superposition of sensor signals when sent over a MAC. The resulting interference gets even worse in emerging paradigms such as the Internet-of-Things (IoT) that involve coexistence of a multitude of sensors transmitting sensed information to a DFC, thereby forming dense WSNs. Moreover MIMO and mMIMO techniques can only enhance diversity gain at the cost of system throughput or vice versa \cite{tarokh1999stsk}.

Space-time coding (STC) techniques like space-time block codes (STBC) \cite{alamouti1998stbc}, Vertical Bell Laboratories Layered Space-Time (V-BLAST) \cite{valenzuela1998vblast} and Linear Dispersion Codes (LDCs) \cite{hassibi_ldc} have been introduced in juxtaposition with MIMO in order to achieve a flexible trade-off between diversity and multiplexing gain, but only at the cost of high encoding and decoding complexities. Moreover, combination of MIMO and STC is capable of providing multiplexing and diversity gains in wideband scenario, though leading to high computational complexities.

Employing STC-aided transmission in a `virtual' MIMO or mMIMO based WSN is not a viable option. WSNs are inherently narrowband applications and are constrained by low computational complexity owing to their low energy budget and battery life of the sensors. Moreover, decoding STC coded transmit vectors at the DFC requires higher system knowledge (like channel parameters, sensor local decisions etc.) in comparison to the uncoded option thereby imposing increased complexity.

The primary contribution of this letter is to propose space-time spreading (STS) of sensor decisions on the transmit side before receiving them over a MAC and fusing the decisions at the DFC with the aim of achieving significant improvement in fusion performance in presence of deep fading and shadowing. STS belongs to a family of coherent shift keying modulation techniques (Spatial Modulation (SM) \cite{yun2008sm}, Space-Shift Keying (SSK) \cite{ceron2009ssk} and Space-Time Shift Keying (STSK) \cite{sugiura2010stsk}) that can champion in offering both diversity and multiplexing gain in a narrowband scenario at reduced complexity. Therefore, the key idea of the STS aided WSN is to encode the local decision of each sensor on a space-time block of fixed duration through the use of dispersion vectors, in order to strike a flexible balance between achievable diversity and multiplexing gain as well as eliminate any chance of ISEI and ISCI.

\section{System Model}
\label{S2}
\subsection{Sensing and Encoding Local Decisions}
We consider a collection of sensors $m \in \mathcal{M} \overset{\triangle}{=} \{1, \dotso, M\}$ deployed to take a local decision autonomously based on a binary hypothesis test, $\mathcal{H}_0$ or $\mathcal{H}_1$, concerning absence and presence of a target of interest respectively. The local decision taken by the $m$th sensor is first mapped to a binary phase shift keyed (BPSK) symbol, $x^l_m \in \mathcal{X} \overset{\triangle}{=} \{+1, - 1\}$ multiplied by a $T$-length dispersion vector, ${\mathbf{a}}^{q, l}_m$, transmitted by the $m$th sensor in the $l$th time-slot to yield, ${\mathbf{s}}^l_m = x^l_m {\mathbf{a}}^{q, l}_m \in \mathbb{C}^{1 \times T}$ for $(l = 1, 2, \dotso, L_f)$. Here, ${\mathbf{a}}^{q, l}_m = [a^{q,l}_{m,1}, a^{q,l}_{m,2}, \dotso, a^{q,l}_{m,T}] \in \mathbb{C}^{1 \times T}$ is the $m$th row of the $q$th space-time dispersion matrix $\mathbf{A}^l_q = [{\mathbf{a}}^{q, l}_1, {\mathbf{a}}^{q, l}_2, \dotso, {\mathbf{a}}^{q, l}_M]^t \in \mathbb{C}^{M \times T}$ selected out of the $Q$ space-time matrices taken from the set ${\{\mathbf{A}^l_q\}}^Q_{q = 1}$ and $L_f$ denotes the total number of space-time blocks in each transmission frame. The encoded set of sensor decisions $\mathbf{S}^l \in \mathbb{C}^{M \times T} \overset{\triangle}{=} \big[{\mathbf{s}}^l_1, {\mathbf{s}}^l_2, \dotso, {\mathbf{s}}^l_M\big]^t = \big[x^l_1{\mathbf{a}}^{q,l}_1, x^l_2{\mathbf{a}}^{q,l}_2, \dotso, x^l_M{\mathbf{a}}^{q,l}_M\big]^t$ must include space-time dispersion matrices that satisfy the power constraint of $tr({\mathbf{A}^l_q}^H\mathbf{A}^l_q) = T~\forall~q$ to ensure unity energy over each space-time block. For ease of representation, we employ parametric system definition in terms of $(M, N, T, Q)$ for any STS-aided WSN. The communication scenario considered for the $l$th time-slot is illustrated in Fig. 1

\begin{figure}[t]
\begin{center}
 \includegraphics[width=8cm, height=8cm]{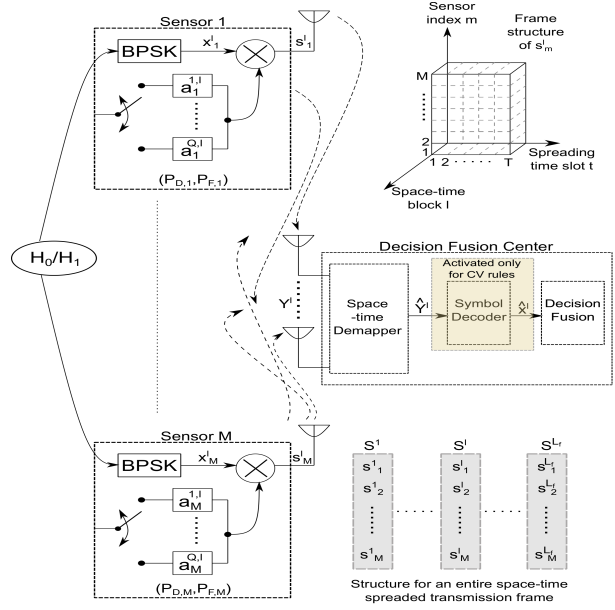}
\end{center}
\caption{Space-time Spreading aided WSN with distributed MIMO DF.}
\label{FIG1}
\end{figure}
\subsection{Signal Model}\label{S2.2}
If the DFC is equipped with $N$ receive antennas, the generic $N \times T$ discrete-time received signal matrix at the DFC is denoted by, $\mathbf{Y}^l \overset{\triangle}{=} \big[{\mathbf{y}}^l_1, {\mathbf{y}}^l_2, \dotso, {\mathbf{y}}^l_N\big]^t = \sqrt{{\rho}^l} \mathbf{G}^l \mathbf{S}^l + \mathbf{w}^l$, where $\mathbf{Y}^l \in \mathbb{C}^{N \times T}$, $\mathbf{G}^l \in \mathbb{C}^{N \times M}$ and $\mathbf{w}^l \sim \mathcal{N}_{\mathbb{C}} \big(\mathbf{0}_{N \times T}, \sigma^2_{w,l}\mathbf{I}_{N \times T}\big)$
\footnotemark[1] are the received signal vector, the channel matrix and the noise vector respectively. The constant $\rho^l$ denotes the energy spent by any of the sensors during the reporting phase and $\mathbf{G}^l$ includes all the samples of the channel impulse response (CIR) between the sensors and the DFC over the $l$th space-time block.

By applying the vectorial stacking operation $vec()$ to the received signal block $\mathbf{Y}^l$ at the space-time de-mapper, we arrive at the linearized equivalent received signal model formulated as, $\hat{\mathbf{Y}}^l = \sqrt{{\rho}^l} \hat{\mathbf{G}}^l \hat{\mathcal{A}}^l \hat{\mathbf{K}}^l + \hat{\mathbf{w}}^l$ where $\hat{\mathbf{Y}}^l = vec(\mathbf{Y}^l) \in \mathbb{C}^{NT \times 1}$, $\hat{\mathbf{G}}^l = \mathbf{I}_g \otimes \mathbf{G}^l \in \mathbb{C}^{NT \times MT}$ (where $\mathbf{I}_g = \mathbf{I}_{T \times T}$) is the estimated channel matrix and $\otimes$ is the Kronecker product, $\hat{\mathbf{w}}^l = vec(\mathbf{w}^l) \in \mathbb{C}^{NT \times 1}$, $\hat{\mathcal{A}}^l = \mathbf{I}_a \otimes \mathbf{A}^l \in \mathbb{C}^{MT \times MQ}$ (where $\mathbf{I}_a = \mathbf{I}_{M \times M}$ and $\mathbf{A}^l \overset{\triangle}{=} \big[{\mathbf{A}}^l_1, {\mathbf{A}}^l_2, \dotso, {\mathbf{A}}^l_Q\big]^t$) and $\hat{\mathbf{K}}^l = vec(\mathbf{K}^l) \overset{\triangle}{=} vec\big([\mathbf{0}, \dotso, \mathbf{0}, \mathbf{x}^l, \mathbf{0}, \dotso, \mathbf{0}]\big)\in \mathbb{C}^{MQ \times 1}$. The transmitted local decision vector $\mathbf{x}^l = [x^l_1, x^l_2, \dotso, x^l_M]^t \in \mathbb{C}^{M \times 1}$ is situated in the $q$th position, noting that the index $q$ corresponds to the index of the dispersion matrix $\textbf{A}^l_q$ activated during $l$th space-time block.

The DFC is in charge of providing a reliable decision (i.e. $\mathcal{H}^1$, \dotso, $\mathcal{H}^{L_f}$) on the basis of the superimposed received decisions taken locally by the sensors independently over each space-time block (i.e. $\hat{\mathbf{Y}}^1, \dotso, \hat{\mathbf{Y}}^{L_f}$). More specifically, we denote $\mathbf{P}^l_{D,m} \overset{\triangle}{=} P\big(\hat{\mathbf{K}}^l_m = [0, \dotso, 0, x^l_m = 1, 0, \dotso, 0]^t|\mathcal{H}_1\big)$ and $\mathbf{P}^l_{F,m} \overset{\triangle}{=} P\big(\hat{\mathbf{K}}^l_m = [0, \dotso, 0, x^l_m = 1, 0, \dotso, 0]^t|\mathcal{H}_0\big)$ respectively, the probabilities of detection and false alarm of the $m$th sensor on the $l$th space-time block.

\footnotetext[1]{The noise vector also accounts for different levels of CSI estimation error, where the estimated channel on the receiver side is contaminated by additive Gaussian noise.}

\subsection{Channel Model}\label{S2.3}

The generic channel coefficient vector $g^l_{n,m}$ is expressed as $g^l_{n,m} = \sqrt{\lambda_m} h^l_{n,m}$, where $\lambda_m$ models the geometric attenuation and shadow fading and remains constant over $n$ and $l$. Based on these assumptions, we have $\mathbf{G}^l = \mathbf{H}^l\sqrt{\mathbf{D}}$ where $\mathbf{G}^l \in \mathbb{C}^{N \times M}$ denotes the matrix of the generic channel coefficients, $\mathbf{H}^l \in \mathbb{C}^{N \times M}$ denotes the matrix of the fast fading coefficients and $\mathbf{D} \in \mathbb{C}^{M \times M}$ is a diagonal matrix with $d_{m,m} = \lambda_m$.

Throughout this letter, we consider that the DFC estimates the CSI, where part of the coherence interval is used for training to estimate the channel and establish the frequency and timing synchronization. We have assumed a coherence interval of $2T$, of which $T$ is the space-time block duration and $T$ is the training phase duration. It is worth mentioning that the linearized sensor-DFC system model contains $M$ non-zero symbol components in $\hat{\mathbf{K}}^l$ given by $\mathbf{x}^l$. Since the equivalent system model is free from the effects of ISCI, we can assume sensing and reporting of each sensor to be independent and decoupled for the formulations done.
\vspace{3mm}

\section {Fusion Rules} \label{S3}
\subsection{Optimum Rule}

The test statistics for energy detector for each space-time block is computed in terms of $\hat{\mathcal{H}}$ and $\mathbf{\Gamma}_{\text{opt}}^l$ denoting the hypotheses and the log-likelihood ratio (LLR) respectively. Exploiting the independence of $\hat{\mathbf{Y}}^l$ from $\mathcal{H}_j^l$, given $\mathbf{x}^l$, an explicit expression of the LLR is obtained as,

\begin{align} \label{eq2}
\mathbf{\Gamma}_{\text{opt}}^l \approx \ln \Bigg[\frac{\sum_{\mathbf{x}^l} \exp\Big(-\frac{||\hat{\mathbf{Y}}^l - \sqrt{\rho^l}(\hat{\mathbf{G}}^l \hat{\mathcal{A}}^l_q) \mathbf{x}^l||^2}{\sigma^2_{w, l}}\Big) P\big(\mathbf{x}^l|\mathcal{H}_1^l\big)}{\sum_{\mathbf{x}^l} \exp\Big(-\frac{||\hat{\mathbf{Y}}^l - \sqrt{\rho^l}(\hat{\mathbf{G}}^l \hat{\mathcal{A}}^l_q) \mathbf{x}^l||^2}{\sigma^2_{w, l}}\Big) P\big(\mathbf{x}^l|\mathcal{H}_0^l\big)}\Bigg]
\end{align}

where $\hat{\mathbf{G}}^l \in \mathbb{C}^{NT \times MT}$, $\hat{\mathcal{A}}^l_q \in \mathbb{C}^{MT \times M}$ and $\mathbf{x}^l \in \mathbb{C}^{M \times 1}$. The expression in (\ref{eq2}) is also numerically unstable due to the presence of exponential functions with large dynamics especially for high SNR and/or large $M$. We will resort to some sub-optimum rules in turn. They are easier to implement, require very little knowledge of the system parameters and offer numerical stability for realistic SNR values.

\subsection{Maximal Ratio Combining (MRC) and modified MRC Rules}

The LLR in (\ref{eq2}) can be simplified under the assumption of perfect sensors as, $P(\mathbf{x}^l = \mathbf{1}_M | \mathcal{H}_1^l) = P(\mathbf{x}^l = - \mathbf{1}_M | \mathcal{H}_0^l) = 1$. In this case, $\mathbf{x}^l \in \{\mathbf{1}_M, - \mathbf{1}_M\}$ and (\ref{eq2}) reduces to,

\begin{align} \label{eq3}
\ln \Bigg[\frac{\exp\Big(-\frac{||\hat{\mathbf{Y}}^l - \sqrt{\rho^l}(\hat{\mathbf{G}}^l \hat{\mathcal{A}}^l_q) \mathbf{1}_M||^2}{\sigma^2_{w, l}}\Big)}{\exp\Big(-\frac{||\hat{\mathbf{Y}}^l + \sqrt{\rho^l}(\hat{\mathbf{G}}^l \hat{\mathcal{A}}^l_q) \mathbf{1}_M||^2}{\sigma^2_{w, l}}\Big)}\Bigg] \propto \mathbb{R}\big\{\big(\mathbf{r}^l_{\text{MRC}}\big)^{\dagger}\hat{\mathbf{Y}}^l\big\}
\end{align}

where, $\mathbb{R}$ represents the real-part of the argument, $\mathbf{\Gamma}^l_{\text{MRC}} = \mathbf{r}^l_{\text{MRC}} \triangleq (\hat{\mathbf{G}}^l \hat{\mathcal{A}}^l_q) \mathbf{1}_M$. It is a sub-optimal rule since in practice, the sensor local decisions are far from being perfect owing to noise, hysteresis error, sensitivity error etc. In order to exploit the linear SNR increase with $N$, we devise the alternative form of MRC, called modified MRC (mMRC) given by, $\mathbf{\Gamma}_{\text{mMRC}}^{l} \triangleq \mathbb{R}\big\{\big(\mathbf{r}^l_{\text{mMRC}}\big)^{\dagger}\hat{\mathbf{Y}}^l\big\}$ where, $\mathbf{r}^l_{\text{mMRC}} \triangleq (\hat{\mathbf{G}}^l \hat{\mathcal{A}}^l_q)((\mathbf{D}_g^l)^{- 1})\mathbf{1}_M$, where $\mathbf{D}_g^l = \frac{1}{N}(\hat{\mathbf{G}}^l \hat{\mathcal{A}}^l_q)^{\dagger}(\hat{\mathbf{G}}^l \hat{\mathcal{A}}^l_q)$ is a diagonal matrix for $N >> M$. It can be observed that mMRC applies a sort of static zero forcing in order to remove dependence on large scale fading coefficients.

\subsection{Widely Linear (WL) Rules}

The test statistics $\mathbf{\Gamma}_{i, l}^{\text{WL}}$ arises from WL processing of $\hat{\mathbf{Y}}^l$, such that $\mathbf{\Gamma}_{i,l}^{\text{WL}} \triangleq (\underline{\mathbf{r}}_{\text{WL},i}^l)^{\dagger} \underline{\hat{\mathbf{Y}}}^l$ and $\underline{\mathbf{r}}_{\text{WL},i}^l$\footnotemark[2] is chosen such that the deflection measure is maximized following, $\underline{\mathbf{r}}^l_{\text{WL}, i} \triangleq \text{max}_{\underline{\mathbf{r}}^l : ||\underline{\mathbf{r}}^l||^2}\mathcal{D}_i (\underline{\mathbf{r}}^l)$, where $\mathcal{D}_i (\underline{\mathbf{r}}^l) \triangleq (\mathbb{E}\{\mathbf{\Gamma}_{l}^{\text{WL}}|\mathcal{H}^l_1\} - \mathbb{E}\{\mathbf{\Gamma}_{l}^{\text{WL}}|\mathcal{H}^l_0\})^2/\mathbb{V}\{\mathbf{\Gamma}_{l}^{\text{WL}}|\mathcal{H}^l_i\}$, $\mathcal{D}_0 (\cdot)$ correspond to the normal and and $\mathcal{D}_1 (\cdot)$ corresponds to the modified deflection \cite{quan2008}. The explicit expressions for $\underline{\mathbf{r}}^l_{\text{WL}, i}$ can be given by,

\begin{align} \label{eq4}
\underline{\mathbf{r}}^l_{\text{WL}, i} = \frac{\mathbf{\Sigma}^{- 1}_{\underline{\hat{\mathbf{Y}}}^l|\hat{\mathbf{G}}^l \hat{\mathcal{A}}^l_q, \mathcal{H}^l_i} \underline{\hat{\mathbf{G}}^l} \hat{\mathcal{A}}^l_q \pmb{\mu}_i^l}{\Big|\Big|\mathbf{\Sigma}^{- 1}_{\underline{\hat{\mathbf{Y}}}^l|\hat{\mathbf{G}}^l \hat{\mathcal{A}}^l_q, \mathcal{H}^l_i} \underline{\hat{\mathbf{G}}^l} \hat{\mathcal{A}}^l_q \pmb{\mu}_i^l\Big|\Big|}
\end{align}

following the proposition made in \cite{ciuonzo2015}, where $\mathbf{\Sigma}^{- 1}_{\underline{\hat{\mathbf{Y}}}^l|\hat{\mathbf{G}}^l \hat{\mathcal{A}}^l_q, \mathcal{H}^l_i} = \big(\rho^l \underline{\hat{\mathbf{G}}^l}\hat{\mathcal{A}}^l_q \mathbf{\Sigma}_{\mathbf{x}^l|\mathcal{H}^l_i} (\underline{\hat{\mathbf{G}}^l}\hat{\mathcal{A}}^l_q)^{\dagger} + \sigma^2_{w, l} \mathbf{I}_{2N}\big)$ and $\pmb{\mu}_i^l \triangleq 2 \big[\big(P_{D, 1}^{l} - P_{F, 1}^{l}\big) \dotso \big(P_{D, M}^{l} - P_{F, M}^{l}\big)\big]^t$.
The aforementioned expressions are based on the fact that the deflection-optimization is optimal only for a mean-shifted Gauss-Gauss hypothesis testing where normal and modified deflections coincide and they both represent the SNR of the statistics under Neyman-Pearson framework.

\footnotetext[2]{$\underline{\mathbf{u}}$ (resp. $\underline{\mathbf{U}}$) denotes the augmented vector (resp. matrix) of $\mathbf{u}$ (resp. $\mathbf{U}$) i.e., $\underline{\mathbf{u}} \triangleq [\mathbf{u}^t~~\mathbf{u}^{\dagger}]^t$ (resp. $\underline{\mathbf{U}} \triangleq [\mathbf{U}^t~~\mathbf{U}^{\dagger}]^t$)}

\subsection{Max-Log Rule}

This sub-optimum fusion rule can be expressed as the difference between hypothesis' prior-weighted minimum distance searches, in the form of,
\begin{align} \label{eq5}
&\mathbf{\Gamma}^l_{\text{Max-Log}} = \overset{\text{min}}{\mathbf{x}^l} \Big[{||\hat{\mathbf{Y}}^l - \sqrt{\rho^l}(\hat{\mathbf{G}}^l \hat{\mathcal{A}}^l_q) \mathbf{x}^l||^2}/{\sigma^2_{w, l}} - \ln P(\mathbf{x}^l|\mathcal{H}_1)\Big] \nonumber\\
&~~- \overset{\text{min}}{\mathbf{x}^l} \Big[{||\hat{\mathbf{Y}}^l - \sqrt{\rho^l}(\hat{\mathbf{G}}^l \hat{\mathcal{A}}^l_q) \mathbf{x}^l||^2}/{\sigma^2_{w, l}} - \ln P(\mathbf{x}^l|\mathcal{H}_0)\Big]
\end{align}
which is an approximation from the turbo codes \cite{Ciuonzo2012} and LLR formulated in this letter.

\subsection{Chair-Varshney (CV) Rules}

If the symbol decoder block at the DFC (refer to Fig.~\ref{FIG1}) computes the estimate $\bar{\mathbf{x}}^l$ of $\mathbf{x}^l$ from $\hat{\mathbf{Y}}^l$, the global decision $\hat{\mathcal{H}}$ is taken on the basis of $\bar{\mathbf{x}}^l$. The Chair-Varshney (CV) rule for noiseless channel is then given by, $\mathbf{\Gamma}^l_{\text{CV}} = \big(\frac{\bar{\mathbf{x}}^l + 1}{2}\big) \ln \big(\frac{P(\mathbf{x}^l|\mathcal{H}_1)}{P(\mathbf{x}^l|\mathcal{H}_0)}\big) + \big(1 - \frac{\bar{\mathbf{x}}^l + 1}{2}\big) \ln \big(\frac{1 - P(\mathbf{x}^l|\mathcal{H}_1)}{1 - P(\mathbf{x}^l|\mathcal{H}_0)}\big)$. Here we consider two different decoders to estimate $\bar{\mathbf{x}}^l$. The first one is the maximum likelihood (ML) detector which can be formulated as, $\bar{\mathbf{x}}^l_{\text{ML}} = \overset{\text{argmin}}{\mathbf{x}^l}{||\hat{\mathbf{Y}}^l - \sqrt{\rho^l}(\hat{\mathbf{G}}^l \hat{\mathcal{A}}^l_q) \mathbf{x}^l||^2}$. The second one is the minimum mean squared error (MMSE) detector expressed as $\bar{\mathbf{x}}^l_{\text{MMSE}} = (\mathbf{r}^l_{\text{MMSE}})^{\dagger}\hat{\mathbf{Y}}^l$ where $\mathbf{r}^l_{\text{MMSE}} \overset{\triangle}{=} \hat{\mathbf{G}}^l \hat{\mathcal{A}}^l_q \Big(\mathbf{D}^l_g + \frac{\sigma^2_{w,l}}{\sqrt{\rho^l}}\mathbf{I}_M\Big)^{-1}$. Once $\bar{\mathbf{x}}^l$ is obtained, we plug it in the CV-rule to obtain the test statistics for CV-ML and CV-MMSE rules.

\section{Performance Analysis}

\subsection{Performance Measures}

Combining the decisions from all the $M$ sensors independently over each space-time block, we can arrive at the total probabilities $P^l_{D_0}$ and $P^l_{F_0}$ for the presented network. Here we compare the performance of different fusion rules both in terms of system false alarm and detection probabilities as, $P^l_{F_0}(\gamma^l, \hat{\mathbf{G}}^l \hat{\mathcal{A}}^l_q) \overset{\triangle}{=}~\text{Pr}\big\{\Gamma^l > \gamma^l|\hat{\mathbf{G}}^l \hat{\mathcal{A}}^l_q, \mathcal{H}^l_0\big\}$ and $P^l_{D_0}(\gamma^l, \hat{\mathbf{G}}^l \hat{\mathcal{A}}^l_q) \overset{\triangle}{=}~\text{Pr}\big\{\Gamma^l > \gamma^l|\hat{\mathbf{G}}^l \hat{\mathcal{A}}^l_q, \mathcal{H}^l_1\big\}$ respectively, where $\Gamma^l$ is the generic test statistics employed at the DFC over the $l$th space-time block and $\gamma^l$ is the threshold with which the test statistics is compared to.
\begin{figure}[t]
\vspace*{-2mm}
\hspace{2cm}
\includegraphics[width=24cm, height=10cm]{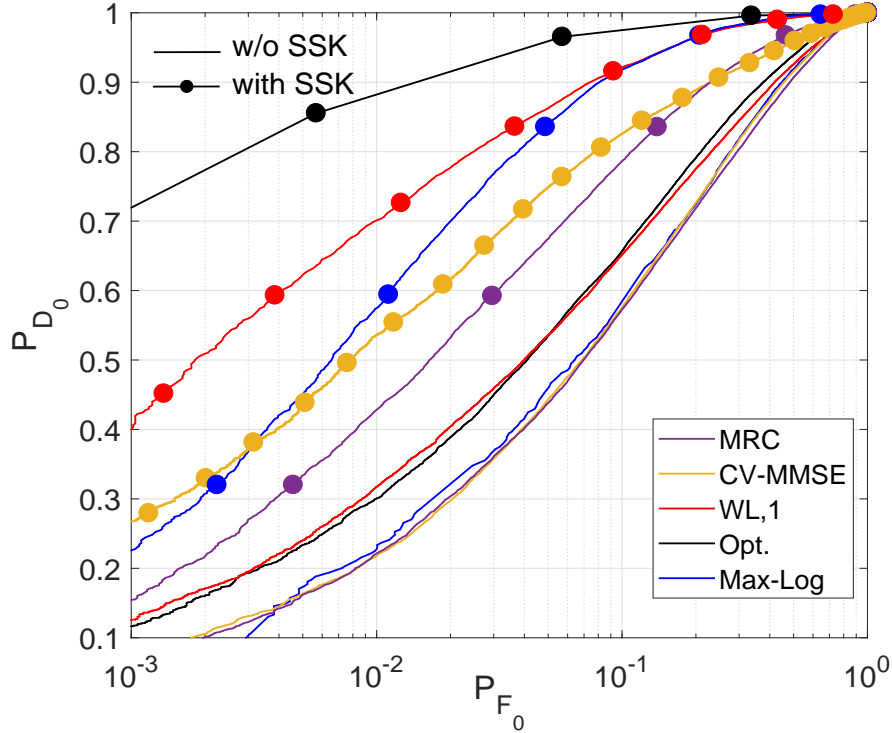}
\caption{Comparative ROC ($P_{D_0}$ v/s $P_{F_0}$) of different fusion rules in $(8, 8, 8, 8)$ STS-aided WSNs with that in WSNs without STS-aided decision transmission over a fixed SNR of 15 dB. Both the sensors and DFC are deployed in a variety of indoor environments.}
\label{FIG3}
\end{figure}
We highlight that $P^l_{D_0}$ should be high to ensure higher probability of correct detection of presence or absence of a target. On the other hand, low values of $P^l_{F_0}$ are necessary to maintain high opportunistic throughput. The choice of threshold $\gamma^l$ leads to a trade-off between $P^l_{F_0}$ and $1 - P_{D_0}^l$.

\subsection{Simulation Set-up}

We simulate performance of a $(M, N, T, Q)$ STS-aided WSN, where $M$ sensors are randomly deployed and uniformly distributed in a circular annulus around the DFC with radius $\phi_{\text{max}} = 1000$ m and $\phi_{\text{min}} = 100$ m. We assume $T = M$ to exploit full diversity order and $Q = M$ to enhance opportunistic throughput of the network. Specifically, ${\{\mathbf{A}^l_q\}}^Q_{q = 1}$, each obeying the power constraint, are randomly generated using Gaussian distribution and the corresponding $P_{D_0}$ for the optimum rule is calculated. The random matrix generation process is repeated 10 times. Out of these, the set of ${\{\mathbf{A}^l_q\}}^Q_{q = 1}$ exhibiting the highest $\{P_{D_0}\}^{\text{Opt.}}$ is chosen for simulating performance. The space-time spreaded local decisions of the sensors are transmitted over a log-normal shadowed ($\lambda_m^l = \psi_m (\frac{\phi_{\text{min}}}{\phi_m})^{\eta}$ where $10\log_{10}(\psi_m) \sim \mathcal{N}(\mu_{\lambda}~\text{dB}, \sigma^2_{\lambda}~\text{dB})$, $\eta$ is the pathloss exponent and $\phi_m$ is the distance of the $m$th sensor to the DFC) and Rayleigh block faded channel ($\mathbf{h}^l_{n,m} \sim \mathcal{N}_{\mathbb{C}}(0, \text{diag}(\mathcal{B}^l_m))$, where $\mathcal{B}^l_m = \big(\beta^l_m(0), \dotso, \beta^l_m(T - 1)\big)^t$ is the channel power delay profile with $\sum_{\tau = 0}^{T - 1} \beta^l_m(\tau) = 1$). We also assume $\rho^l = 1/\sqrt{N}$ and independently and identically distributed (iid) decisions with $({P}^l_{D}, {P}^l_{F}) = (0.5, 0.05)$.
\begin{figure}[t]
\vspace*{-2mm}
\hspace{2cm}
 \includegraphics[width=24cm, height=10cm]{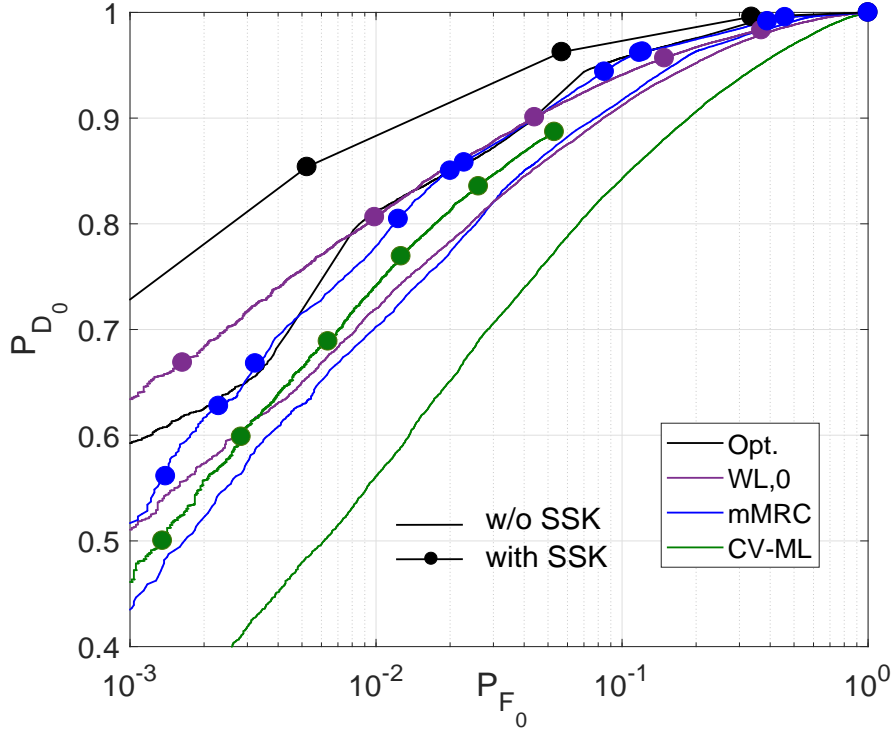}
\vspace*{-6mm}
\caption{Comparative ROC ($P_{D_0}$ v/s $P_{F_0}$) of different fusion rules in $(10, 100, 10, 10)$ STS-aided WSNs with that in WSNs without STS-aided decision transmission over a fixed SNR of 15 dB. Both the sensors and DFC are deployed in a variety of indoor environments.}
\label{FIG4}
\end{figure}

\subsection{Simulation Results}

In Fig.~\ref{FIG3} and Fig.~\ref{FIG4}, we present the ROC of all the fusion rules for two different configurations of WSNs, a) fully-loaded MIMO set-up ($M = 8$, $N = 8$) and b) virtual mMIMO set-up ($M = 10$, $N = 100$). We simulate performance of the formulated fusion rules over a MAC with pathloss exponent, $\eta$ of 2, experiencing moderate shadowing, $(\mu_{\lambda}, \sigma_{\lambda}) = (4, 2)$ dB. The above-mentioned parameters are representative of a variety of indoor environments. For the two network set-ups, STS aided sensor decision transmission offers significant improvement in performance over that without STS. Each fusion rule gain in performance from a minimum of 3 times (MRC) to a maximum of 6 times (Opt.) in case of fully-loaded MIMO, and a maximum of $9/8$ times (Opt.) to a minimum of $4/3$ times (CV-ML) in case of virtual mMIMO set-up. For both the set-ups, MRC (mMRC for mMIMO case) and CV-ML performs worst respectively, as corroborated in \cite{ciuonzo2015,Ciuonzo2012}.
\begin{figure}[t]
\vspace*{-2mm}
\begin{center}
\hspace*{2cm}
 \includegraphics[width=24cm, height=10cm]{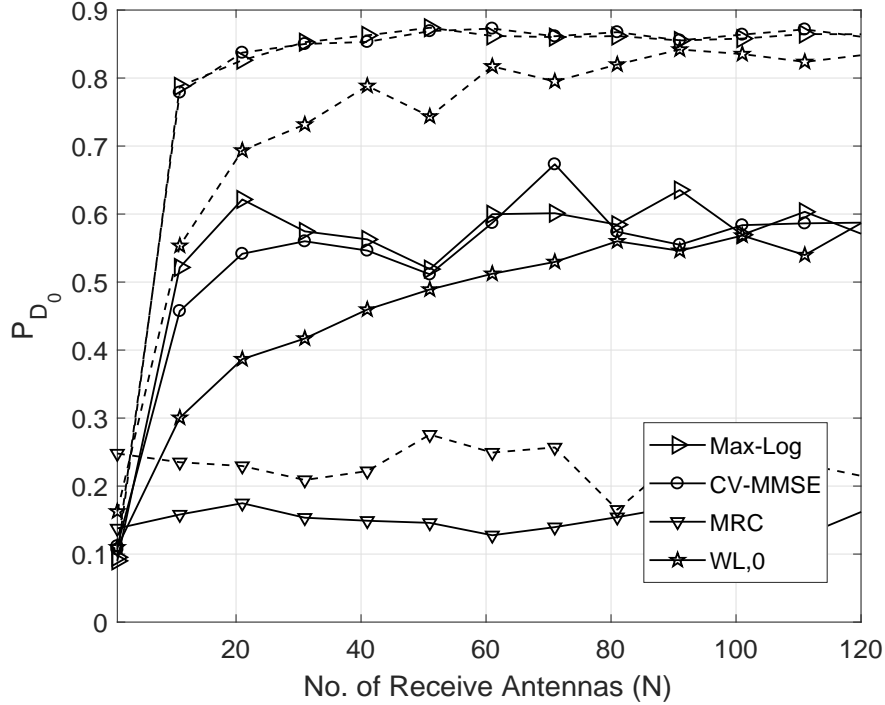}
\end{center}
\caption{Variation in probability of detection of different fusion rules with changing $N$ ($P_{D_0}$ v/s $N$) for two cases, $(4, N, 8, 8)$ (solid line) and $(8, N, 8, 8)$ (dashed line) STS-aided WSNs over a fixed SNR of 15 dB and $P^l_{F_0} = 0.01$. The sensors are deployed in a tunnel-like environment.}
\label{FIG5}
\end{figure}
In Fig.~\ref{FIG5}, we plot $P_{D_0}$ of the presented fusion rules as a function of $N$ under $P_{F_0} \leq 0.01$; we depict the cases $M \in \{4, 8\}$ and $Q = T = M$ for each case. Performance of all fusion rules improves with the increase in $N$, however reaches saturation depending on the SNR and the chosen fusion rule. Some rules like CV-MMSE, and Max-Log ($N > 40$) proceeds to saturation faster than other rules like WL,0 ($N > 90$). It is also evident that MRC performs worse than any other fusion rule, as MRC does not exploit STS aided local sensor performance at the decoding stage like WL, CV-MMSE or Max-Log. Indeed, the probability of detection with MRC is dependent only on the channel statistics.

\begin{figure}[t]
\vspace*{-5mm}
\begin{center}
\hspace*{2cm}
 \includegraphics[width=24cm, height=10cm]{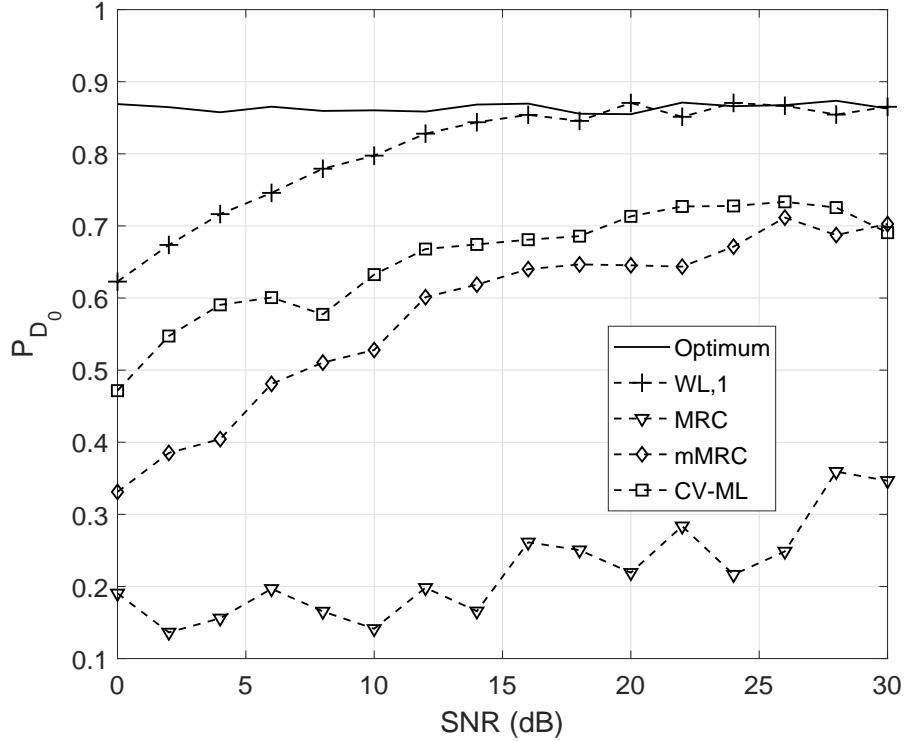}
\end{center}
\caption{Comparative probability of detection performance ($P_{D_0}$ v/s SNR(dB)) for different fusion rules in an $(8, 32, 8, 8)$ STS-aided WSN where outdoor sensors communicate with an indoor DFC.}
\label{FIG6}
\end{figure}
In Fig.~\ref{FIG6}, we demonstrate $P_{D_0}$ of the presented fusion rules as a function of $(SNR)_{dB}$, where the SNR measure includes both the channel noise and varying levels of CSI estimation errors. In this case, $N = 32, Q = T = 8$ and $\eta  = 5$ and $\mu_{\lambda} = 4$ dB, a representative condition of outdoor sensors communicating with indoor DFC. CV-MMSE, Max-Log and WL,1 rules approach the optimal performance at moderate to high SNRs. However, MRC, mMRC and CV-ML rules fail to achieve optimal performance even at high SNRs, as opposed to the observations made in \cite{dey2019tap}. It has been demonstrated in \cite{sugiura2010stsk}, that for $T>1$, diversity increases but at the cost of reliability for STSK modulated systems. For $Q>1$, throughput increases but at the cost of degraded bit error rate (BER). Here, we have chosen $T = Q = M$ for STS thereby sacrificing reliability of system knowledge (like CSI statistics, statistics of sensor decision vectors) and lower probability of error for the sake of gain in diversity and network throughput. It can be broadly concluded that CV-ML performs poorly in any network scenario and propagation condition as the CV-ML statistics is dependent on the channel SNR which is kept fixed for Figs.~\ref{FIG3}, \ref{FIG4}, and \ref{FIG5}.

\section{Conclusion}
Inspired by the recent success of STSK scheme in striking a flexible balance between diversity and multiplexing gain, we conceive the novel idea of space-time spreading the local sensor decisions before transmission in a WSN. The resultant network will not only benefit from low-complexity diversity gain and improvement in opportunistic throughput but also from ISCI and ISEI free transmission in a densely deployed scenario. The STS scheme used can be modified depending on the chosen $Q$ and $T$ to include SM, SSK and STSK arrangements. However, using different values for $Q, T, M$ will involve multiple information symbols for carrying sensing decisions, an interesting generalization which we leave for our future work. Our presented simulation results demonstrate the potential of STS-aided WSN in outperforming the conventional MIMO and mMIMO based WSN arrangements. Motivated by this observation, in future, we plan to extend our results under different conditions of dispersion matrix optimization, multi-slot decision transmission and correlated sensor observations in sensing performance.

\bibliographystyle{unsrt}
\bibliography{bibliography}
\end{document}